\documentclass[preprint,superscriptaddress,showpacs]{revtex4-1}
\usepackage{CJKutf8} 
\usepackage{amsmath,amssymb} 
\usepackage{bm}
\usepackage{hyperref}
\usepackage{setspace}

\usepackage{subfigure}
\usepackage{graphicx}
\usepackage{color}
\usepackage{lipsum}

\def\urlprefix{}
\begin{document}

\begin{CJK*}{UTF8}{} 

\title{Observation of ultracold atomic bubbles in orbital microgravity}

\author{R.A.~Carollo$^=$}
\affiliation{Department of Physics and Astronomy, Bates College, Lewiston, ME 04240, USA}

\author{D.C.~Aveline$^=$}
\affiliation{Jet Propulsion Laboratory, California Institute of Technology, Pasadena, CA 91109, USA}

\author{B.~Rhyno} 
\affiliation{Department of Physics, University of Illinois at Urbana-Champaign, Urbana, Illinois 61801-3080, USA}

\author{S.~Vishveshwara}
\affiliation{Department of Physics, University of Illinois at Urbana-Champaign, Urbana, Illinois 61801-3080, USA}

\author{C.~Lannert}
\affiliation{Department of Physics, Smith College, Northampton, Massachusetts 01063, USA}
\affiliation{Department of Physics, University of Massachusetts, Amherst, Massachusetts 01003-9300, USA
}

\author{J.D.~Murphree} 
\affiliation{Department of Physics and Astronomy, Bates College, Lewiston, ME 04240, USA}

\author{E.R.~Elliott} 
\affiliation{Jet Propulsion Laboratory, California Institute of Technology, Pasadena, CA 91109, USA}

\author{J.R.~Williams} 
\affiliation{Jet Propulsion Laboratory, California Institute of Technology, Pasadena, CA 91109, USA}

\author{R.J.~Thompson} 
\affiliation{Jet Propulsion Laboratory, California Institute of Technology, Pasadena, CA 91109, USA}


\author{N.~Lundblad}
\email{nlundbla@bates.edu}
\affiliation{Department of Physics and Astronomy, Bates College, Lewiston, ME 04240, USA}

\date{\today}
\maketitle
\end{CJK*}

{\bf Significant leaps in the understanding of quantum systems have been driven by the exploration of geometry, topology, dimensionality, and interactions with ultracold atomic ensembles~\cite{Hadzibabic:2006lr,Clade.2009,Kinoshita.2004,Eckel:2014gf,Chin.2010}.  A system where atoms evolve while confined on an ellipsoidal surface represents a heretofore unexplored geometry and topology.  Realizing such an ultracold bubble system---potentially Bose-Einstein condensed---has areas of interest including quantized-vortex flow respecting topological constraints imposed by closed surfaces, new collective modes, and self-interference via free bubble expansion~\cite{Tononi.2021,Tononi.2020,Tononi:2019ci,Padavic:2020fj,Sun:2018de,Padavic:2017cv,Lannert:2007kk,Moller:2020ik,Bereta.2021lkc,Zhang:2018gl}.  Large ultracold bubbles, created by inflating smaller condensates, directly tie into Hubble-analog expansion physics~\cite{Banik.2021,Bhardwaj.2021,Eckel:2018gv}.
Here, we report observations from the NASA Cold Atom Lab facility~\cite{Aveline:2020gk} aboard the International Space Station of bubbles of ultracold atoms created using a radiofrequency-dressing protocol.  We observe a variety of bubble configurations of differing sizes and initial temperature, and explore bubble thermodynamics, demonstrating significant cooling associated with inflation.  Additionally, we achieve partial coverings of bubble traps greater than 1 mm in size with ultracold films of inferred few-$\mu$m thickness, and we observe the dynamics of shell structures projected into free-evolving harmonic confinement.  The observations are part of the first generation of scientific measurements made with ultracold atoms in space, exploiting the benefits of perpetual free-fall to explore gravity-free evolution of quantum systems that are prohibitively difficult to create on Earth. This work points the way to experiments focused on the nature of the Bose-Einstein condensed bubble, the character of its excitations, and the role of topology in its evolution; it also ushers in an era of orbital microgravity quantum-gas physics.    }


While the techniques for the generation of ultracold atomic bubbles have been known since 2001~\cite{PhysRevLett.86.1195},  terrestrial gravity prevents the observation of these configurations, as the trapped sample simply sags to the lower fraction of the given shell trap, forming a conventional (if distorted) ultracold ensemble.  With the recent construction of the NASA Cold Atom Lab (CAL) facility and its subsequent delivery to the International Space Station and commissioning as an orbital BEC facility~\cite{Aveline:2020gk,Elliott:2018dk}, experimental efforts requiring a sustained  microgravity environment are now possible, including realistic possibilities for ultracold bubble physics, as recently proposed~\cite{Lundblad:2019ia}.  In what follows, we present observations of ultracold bubbles created in microgravity aboard CAL using protocols developed to explore bubble size and temperature.  We give detailed measurements of subsequent inflated bubble temperature varying as a function of initial sample temperature---linking to theory realistically modelling the CAL apparatus---and observe the effects of shell-trap removal and the resulting atomic bubble propagation in the preexisting harmonic trap.

\begin{figure*}[t]
\centering

\includegraphics[width=\columnwidth]{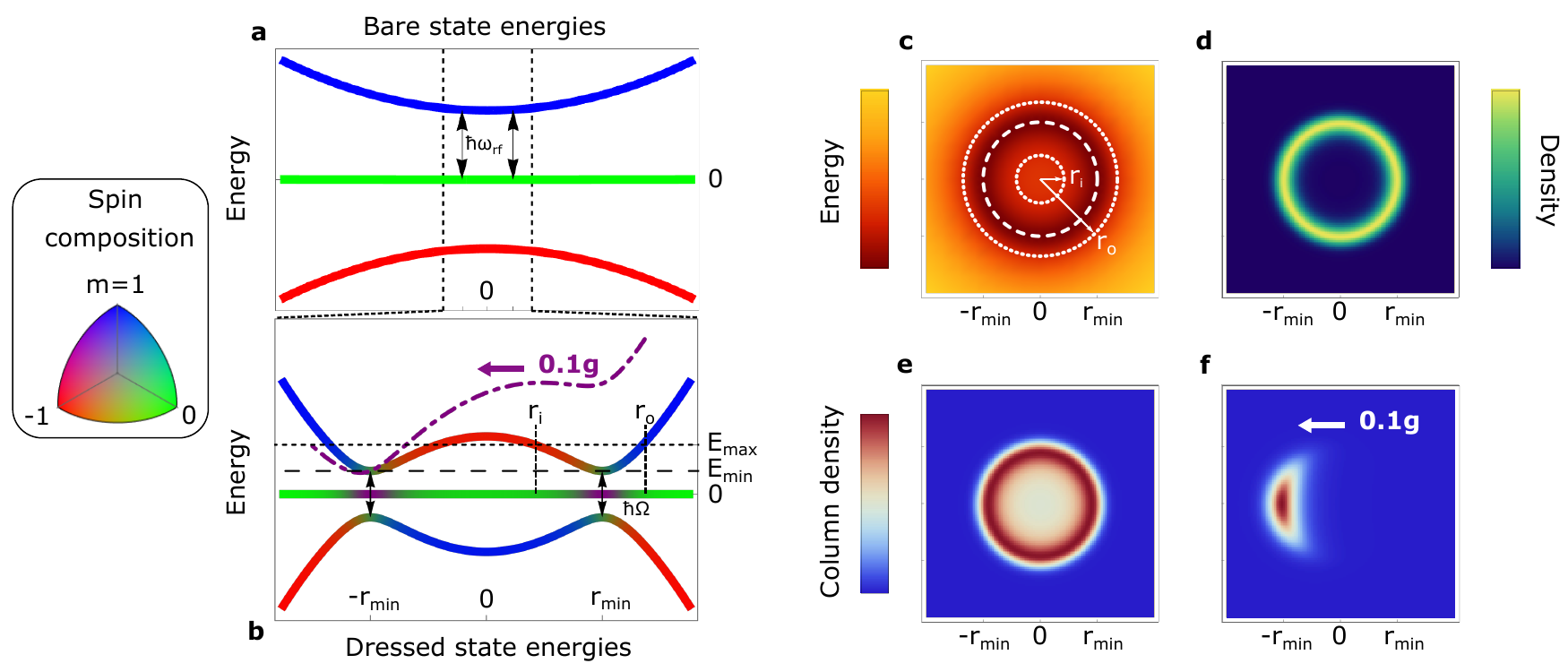}\caption{
\textbf{Creating ultracold bubbles.}  Illustrated by a three-level analytic model of an isotropic trap, using otherwise typical experimental parameters. {\bf a,} Atoms are prepared in the highest energy spin state near the minimum of a static magnetic field
(blue curve). Applying a radio frequency (rf) magnetic field of frequency $\omega_{\rm rf}$ and coupling strength $\Omega$ creates spatially-varying superpositions of the bare magnetic states. {\bf b,} Atoms in stationary (dressed) states of the combined fields experience a dressed potential with extrema at points $\pm \textrm{r}_{\textrm{min}}$ where the rf field is resonant.
The highest-energy dressed potential forms a double well along any axis passing through the static field's minimum; the dot-dashed purple curve shows the effect of adding a gravitational field of magnitude $0.1\, g$.  {\bf c,} Equipotential surfaces in two dimensions, revealing the characteristic bubble shape. Emphasized surfaces---also shown in {\bf b}---are the trap minimum ($E_{\rm min}$ at ${\rm r}_{\rm min}$, dashed) and qualitative bubble boundaries, where the density of a 200 nK thermal cloud drops below 1\% of its maximum ($E_{\rm max}$ at ${\rm r}_{\rm i}$ and ${\rm r}_{\rm o}$, dotted). Atoms congregate at the trap minimum as seen in {\bf d}, the cross-section of the resulting atomic density, and {\bf e}, the column density or optical depth, the latter measured in experiments via absorption imaging. {\bf f}, Column density of atoms in the potential described by the purple curve in {\bf b}, showing that a $0.1\, g$ gravitational field prevents atoms from covering the bubble's entire surface; $1\,g$ creates even greater deformation.}  
\label{dressing}
\end{figure*}

We first summarize the atomic-physics framework for generation of ultracold bubble systems.  Our creation of a shell-like confining potential $U({\bf r})$ for ultracold atoms stems from a theoretical proposal to generate matter-wave bubbles allowing for the study of 2D BECs tightly pinned to partial coverings of the potential~\cite{PhysRevLett.86.1195,Zobay:2000tk,zobay:023605}.  This scheme relies on a locally harmonic spin-dependent trapping potential originating in an applied  magnetic field ${\bf B}({\bf r})$, combined with a near-resonant oscillatory  magnetic field ${\bf B}_{\rm rf}$ at radio frequency $\omega_{\rm rf}$, resulting in spatially-dependent dressed atomic states~\cite{perrin2017trapping,Garraway:2016hu}.  Atoms in these states experience effective (dressed, or adiabatic) potentials which can be tailored such that atoms enter bubble-like configurations of diverse size and thickness.  As depicted in Fig.~\ref{dressing}, atoms in spin states $|m\rangle$  exposed to a magnetic trapping field experience trapping potentials $U_m({\bf r}) = g_F m \mu_B |{\bf B}({\bf r})|$, where $g_F$ is the Land\'{e} g-factor associated with (in our case) a given atomic hyperfine manifold of total angular momentum $F$.    In the presence of ${\bf B}_{\rm rf}$, the combination of a rotating-frame transformation and the application of a rotating-wave approximation results in an dressed-picture Hamiltonian ${\mathcal H}$, and the creation of associated dressed potentials $U^*_{m'}({\bf r})$  (see Methods).  While shell potentials for ultracold atoms have been generated and explored in several groups~\cite{white:023616,Merloti:2013ft,perrinbec}, efforts to explore bubble-centered physics have been hampered by the presence of terrestrial gravitational potential energy $U_g = M g z$.   Preliminary schemes have been developed to cancel this gravitational tilt---using (for example) an appropriate ac Stark shift gradient or rf coupling gradient---however, precise cancellation over a volume appropriate for an ultracold bubble is not yet possible~\cite{Shibata:2020fx,Guo.2021}.   Application of this technique to the CAL apparatus was recently proposed, specifically accounting for known inhomogeneities in realistic dressed potentials such as would originate from the spatial dependence of the radiofrequency  field or the anharmonicity of the CAL magnetic trap~\cite{Lundblad:2019ia}.  While an idealized bubble is spherical~\cite{Meister:2019bv}, these shell potentials are generally ellipsoidal, as dictated by the aspect ratio of the generating trap.      


We conducted these experiments in a remotely-operated user facility located in low Earth orbit aboard the International Space Station (ISS).   This facility, the NASA Cold Atom Lab (CAL), was delivered to ISS via rocket launch in 2018 and conducted its first science runs through January 2020 before undergoing hardware upgrades. Its development and ground test process has been reported~\cite{Elliott:2018dk}, as has its core functionality, the generation of BECs in orbital microgravity~\cite{Aveline:2020gk}.  The regular operation of the facility provides ultracold samples to scheduled users; for this work, typically ensembles of $\sim$10$^4$ $^{87}$Rb atoms at or below the BEC transition temperature $T_c$ were provided in a tightly-confining ``atom chip"-style magnetic trap, although significantly hotter samples were also used.  The facility-provided default trap, common to all users, was not suitable for shell-potential exploration due to its high aspect ratio ($\sim$10) and proximity to the atom-chip surface---effectively the wall of the vacuum chamber.  We thus initiated all experiments with an expansion trajectory designed to bring the ultracold sample away from the vacuum wall and reduce the aspect ratio of its confining trap~\cite{Sackett:2017gp,Pollard:2020gc}. The resulting trap configuration served as an initial condition for these experiments, featuring an ensemble of ultracold $^{87}$Rb atoms, nominally in the $|F=2, m=2 \rangle$ internal state, confined in a trapping potential $U_2({\bf r})$ approximately 700 $\mu$m from the surface of the CAL atom chip.  The trap is described by an aspect ratio of $\sim$3 and a geometric mean trapping frequency of $\overline{\omega} =2\pi\times$ 67(1) Hz (see Methods).   Turning on the coupling radiofrequency field (linearly polarized along $z$, the axis perpendicular to the atom chip) far below resonance projects the system into the appropriate dressed-state manifold---where the dressed state is nearly identical to the initial bare state---with further dynamic alteration occurring via ramps of $\omega_{\rm rf}$.  Typically the frequency is referenced (via a detuning $\Delta = \omega_{\rm rf}-\omega_0$) to an experimentally determined ``trap bottom" defined such that $\hbar\omega_0 = E_{2,2}-E_{2,1}$, namely, the energy separation of the two topmost energy levels in the $F=2$ ground state manifold.  To move to a shell potential of chosen size, the value of $\omega_{\rm rf}$ is linearly ramped at a rate (typically $\sim$1 kHz/ms) chosen for mechanical adiabaticity; see Methods. After rapid ($\sim$ 1--10 $\mu$s) switchoff of both rf field and magnetic trap, imaging of the resulting clouds is performed via destructive absorption imaging.   The parameter space of the resulting datasets is spanned by variation of initial temperature $T$, atom number $N$, final detuning $\Delta$, and the time-of-flight (TOF) between trap snap-off and imaging.  While the rf coupling strength ($\Omega \propto B_{\rm rf}$) can also alter dressed-state trap geometry, for these experiments it was held constant at a value $\Omega_0 = 2\pi\times$ 6(1) kHz, calibrated via Rabi spectroscopy of the atomic clouds.  

\begin{figure*}[t!]
\centering
  \includegraphics[width=\columnwidth]{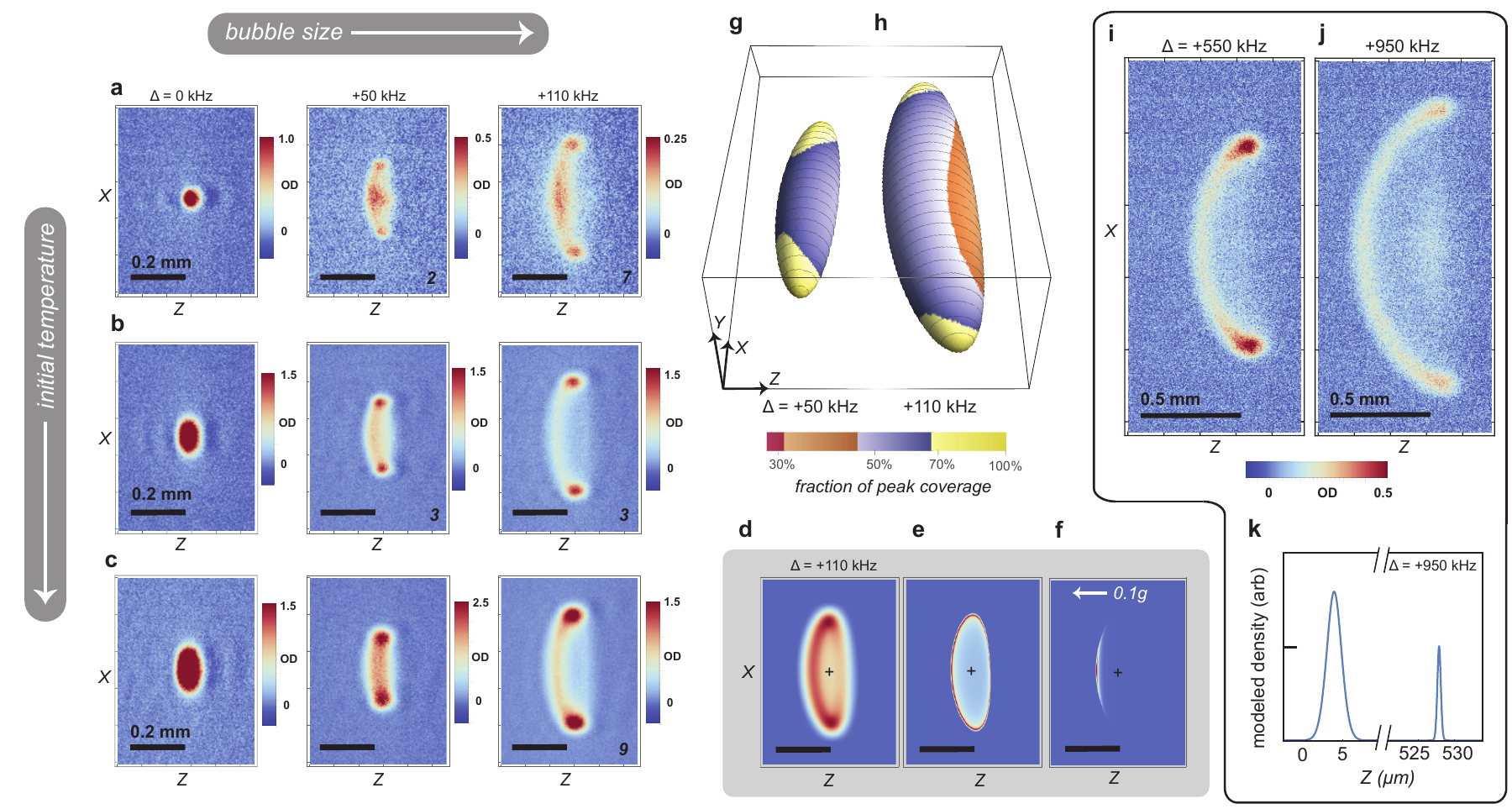}\caption{\textbf{Ultracold bubble observations and modeling.}  {\bf a)} Bubble inflation sequence with  an initial temperature of $\simeq$100 nK (partially condensed) set by rf-knife value near 4.87 MHz; trap sizes given by $\Delta$ parameters of 0 kHz, +50 kHz, and +110 kHz, respectively, left to right.   All data show optical depth (OD); images were taken with minimal  time-of-flight expansion.  {\bf b}, Inflation sequence with initial temperature ($\simeq$300 nK) set by rf-knife value near 4.9 MHz.  {\bf c}, Inflation with initial temperature ($\simeq$400 nK) set by rf-knife value near 4.99 MHz.  In all inflated clouds, note terrestrially-unattainable lobes at $\pm x$.  When present, the number at lower right in a pane denotes the number of images averaged together, originating in identical experimental sequences.   {\bf d,} Model prediction of the $\Delta=+110$ kHz column density at $T_{\rm bubble}=$ 100 nK, akin to the corresponding bubbles in (b-c), where the model includes simple blurring by a point-spread function of width 40 $\mu$m.  {\bf e,} the corresponding non-blurred model column density. {\bf f}, The model predictions of (e) modified by the presence of 10\% of terrestrial gravity, demonstrating the impact of the microgravity environment.  {\bf g-h,} Illustrative model of bubble coverage for $\Delta=+50$ kHz and +110 kHz, both at typical bubble temperatures of 100 nK, showing an approximate factor-of-two variation around the bubble.  Note increased inhomogeneity for the larger bubble, corresponding to residual potential tilts $\sim .005\, g$. {\bf i-j}, Extreme inflation to mm-scale sizes with $\Delta=$ +550 kHz and +950 kHz; initial temperature ($\sim$1 $\mu$K) set by rf-knife value of 5.3 MHz.   {\bf k,} Model prediction of $\Delta=+$950 kHz ensemble at $T_{\rm bubble}=$100 nK; shown is a 1D slice along $z$ of the predicted atomic density distribution. Note $\sim$500:1 ratio of bubble diameter to thickness; also note that while bubble coverage is suppressed at this $\Delta$, it remains discernible.  }     
\label{zoo}
\end{figure*}

Fig.~\ref{zoo} shows a variety of ultracold shell structures we have formed aboard CAL, including predictions of a semiclassical model whose potentials were initially developed in~Ref.~\cite{Lundblad:2019ia}.   All images are column densities, typical of absorption imaging, thus all features are distorted by imaging resolution effects and the effects of shell-trap inhomogeneity; we note that predicted thicknesses of either condensate or thermal shell clouds  in these systems are in the range $\sim$1--10 $\mu\textrm{m}$ as illustrated in in Fig.~2(e,k), revealing the ultracold-atom coverings of these bubble potentials to be remarkably delicate structures, impossible to generate in the presence of terrestrial gravity.  For moderately-sized bubbles as depicted in Fig.~2(a-c) and modeled in Fig.~2(d-h) the modeled thickness of the ultracold atomic film varies by less than a factor of two around the $\Delta=50$ kHz shell, and by a factor of three around the $\Delta=110$ kHz shell.  In the limit of large $\Delta$, shells of diameter at the few-mm scale are possible, as shown in Fig.~2(i-j). While the lobe structures seen in many images at the $\pm x$ ends of the observed clouds are qualitatively observed in modeling through approximate imaging-resolution estimates, as shown in Fig.~2(d), at larger radii this simple modeling does not suffice; a more sophisticated imaging analysis might yield deeper understanding here~\cite{Putra:2014fs}. Residual potential-energy inhomogeneities in the shell potential associated with i) the decrease of the coupling rf amplitude with increasing distance from the antenna, ii) variation of the trap magnetic field direction, and iii) the anharmonicity of the atom-chip magnetic trap, are $\sim$$h\times$100 Hz ($k_B\times 5$ nK) for $\sim$100-$\mu$m-size clouds, corresponding to effective gravitational effects of $\lesssim 0.005\, g$; as such, residual $\mu$g accelerations of the ISS should not be relevant here.       

\begin{figure}[t]
\centering
  \includegraphics[width=6in]{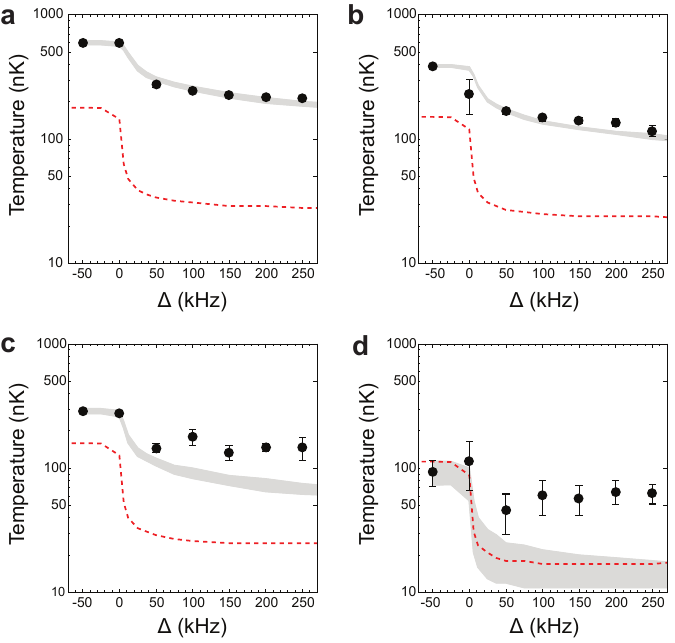}\caption{
\textbf{Thermometry of bubbles.} Shown above are data $T_{\rm bubble}(\Delta)$ (black), where the detuning $\Delta$ serves as a proxy for bubble size; error bars (where visible) represent standard errors. Also shown are theoretical $T_{\rm bubble}(\Delta)$ (gray) and $T_c(\Delta)$ (dashed) predictions for initial (pre-inflation) temperatures set by evaporative-cooling `rf-knife' values, as follows: {\bf a,} 600(20) nK (5.1 MHz) {\bf b,} 390(10) nK (5.0 MHz) {\bf c,} 290(10) nK (4.93 MHz), and {\bf d,} 90(20) nK (4.855 MHz) where the last initial condition is a partial Bose-Einstein condensate, although clouds appear thermal for all positive $\Delta$. Theory curves for $T_c(\Delta)$ are shown for illustrative purposes and assume a typical mean atom numbers of (200(10), 120(5), 140(5), 50(5))$\times 10^3$ for (a)--(d), respectively.  The data show significant cooling as the bubble trap is inflated from an unperturbed initial harmonic trap.}     
\label{thermo}
\end{figure}

In Fig.~\ref{thermo} we show the results of bubble thermometry with associated theoretical modeling of $T_{\rm bubble}(\Delta)$. In order to provide a visual reference for the temperature relative to Bose-Einstein condensation, we also show $T_c(\Delta)$ given typical values of atom number $N$ for a given dataset.  Thermometry is performed through turning off the trapping potential (rf and chip magnetic fields) and letting the cloud expand in time of flight (TOF) up to 48 ms, during which the atoms remain roughly centered around their original location given the weightless environment.  The absorption profile (column density) of the cloud is summed along the $x$ direction and then fit to a Gaussian profile, which while inappropriate for short TOF (and for partially condensed samples) yields a generally correct impression of the long-TOF expansion speed of the released cloud.  The key intuition for the thermodynamics of shell potentials is that the reshaping of the bare magnetic trap into a bubble trap of given radius is equivalent to an adiabatic expansion, albeit one not necessarily proceeding at constant phase space density~\cite{Pinkse:1997cx}. We show thermometry curves for samples initially partially condensed (BEC fraction typically 60\% or higher) (Fig.~4(d)) as well as for samples with initial temperatures up to $> 3 T_c$, shown in Fig.~4(a-c). For all four initial sample temperatures, we observe drastic drops in temperature as bubble size is increased, with the most rapid change occurring over the range of $\Delta$ associated with the atomic cloud hollowing out (as the trapping potential changes from harmonic to shell-like).     

To model these data, we developed estimates for temperatures of ultracold shells using a semiclassical fixed-entropy approach, with the entropy associated with a given theory curve in Fig.~\ref{thermo} set by the initial temperature and number in the given configuration (see Methods). While this modeling approach for $T_{\rm bubble}(\Delta)$ yields good agreement for the hottest initial sample in Fig.~\ref{thermo}(a), the data increasingly show suppressed cooling effects at lower initial temperatures, despite directional agreement.  We attribute these discrepancies, most significantly shown in Fig.~\ref{thermo}(d), to a combination of several possible factors.  On the experimental side this could stem from non-adiabaticity of the inflation process and/or potential systematic errors in thermometry at low temperatures (and low atom number); on the theoretical side, there could be effects associated with the failure of the semiclassical approximation associated with the transition to quasi-2D confinement.

A key feature of bubble thermodynamics is that while calculated and observed $T_{\rm bubble}(\Delta)$ drop precipitously as the trapping potential is adiabatically `inflated', the associated $T_c$ does not drop commensurately. This is caused by an initial drop in phase space density even at constant entropy; this decoupling of phase space density and entropy due to geometrical changes has been exploited in various cold-atom experiments~\cite{Lin:2009p1534,StamperKurn:1998jb,2003Sci...299..232W} but in shell geometry presents an added challenge.  Thus, we find in principle that an initially barely-condensed cloud (such as used in Fig.~\ref{thermo}(d)) should enter the normal phase again upon inflation, even given perfect adiabaticity, and potentially re-condense upon extreme inflation.  This issue (and the thermodynamics of shell inflation in general, including the nature, role, and limits of the semiclassical approximation) is discussed comprehensively in Ref.~\cite{Rhyno.2021}.

\begin{figure}[t]
\centering
  \includegraphics[width=6in]{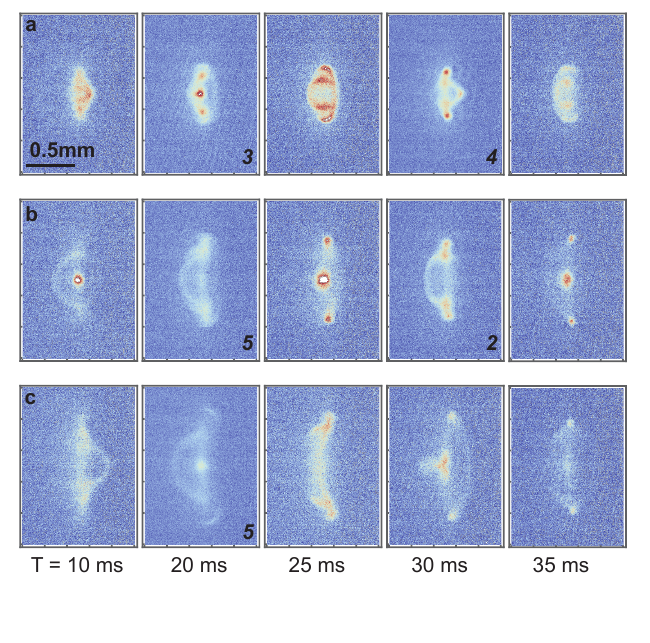}\caption{\textbf{Evolution upon removal of dressing} Rapidly turning off the rf field forces evolution in bare magnetic trap according to projected spin component.  Rows correspond to three different shell sizes: {\bf a,} dressing detuning  $\Delta=150$ kHz,  {\bf b,} $\Delta = 350 $ kHz, and {\bf c,} $\Delta = 550$ kHz.  Evolution time $T$ increases rightward as denoted at bottom. When present, the number at lower right of each image denotes the number of images averaged together, originating in identical experimental sequences.  Note qualitative recurrence timescale of 10 ms, roughly corresponding to the trap oscillation period in the horizontal direction.  
 }     
\label{other}
\end{figure}

Given a dressed (spin-superposition) ultracold shell system, an immediate point of curiosity arises regarding what might happen upon removal of the dressing field while preserving the confining magnetic trap.   Such an action should (in the limit of rapid turn-off) project the dressed bubble eigenstate into its bare spin components, which would then experience the original magnetic trap as dictated by the magnetic moment of those components.  Thus, we would expect an inward-propagating shell to appear as the hold time $T$ in the ``de-dressed" trap is varied.   In Fig.~\ref{other} we show example observations of such propagation of (likely thermal) shell ensembles.   Understanding of the qualitative nature of this effect is an important prologue to understanding the behavior of dressed condensates undergoing similar propagation, which should result in complex interference patterns given by time-evolution of the bare ground-state spin components~\cite{Lannert:2007kk}; it also suggests future investigations along the lines of the ``Bose-nova" collapse experiments~\cite{PhysRevLett.86.4211}.

We have observed and characterized ultracold bubble systems and established a model and theoretical framework for them.  The capacity to perform these experiments is currently unique to laboratories operating in a microgravity environment, and our observations point the way to future work exploring the fundamental nature of the condensed bubble state.  With significantly lower initial temperatures in future experiments, adiabatic inflation would not provoke such significant loss of condensate fraction.  Technical improvements to the experimental hardware and software aboard CAL (as recently implemented in a hardware upgrade, including a larger rf antenna with associated increase in dressing homogeneity) should also improve bubble quality, as could the use of compensatory microwave dressing~\cite{2019PhRvA.100e3416S,Fancher:2018hb}.  Alternatively, planned facilities such as BECCAL \cite{Frye:2021dz} could incorporate secondary evaporative cooling of the dressed clouds~\cite{GarridoAlzar:2006ik}, permitting a direct path to higher condensate fraction. 

Future work (on CAL or successors) could generate vortices in condensate bubbles either through direct stirring or rotation of the dressed trap, or through spontaneous generation of vortices across the condensate phase transition through the Kibble-Zurek mechanism.  Experimental exploration of recent theoretical work regarding the role of the Berezinskii-Kosterlitz-Thouless transition in 2D superfluid bubbles would be a compelling target as a case of the general problem of quantum-gas physics on curved manifolds~\cite{Tononi.2020,Tononi.2021}.   Additionally, multi-axis imaging for complete characterization of the bubble structure should be possible, and implementation of multi-rf-frequency protocols for nested (tunneling) shells is within sight~\cite{Luksch:2019be,Harte:2018jm}, as are experiments aiming at observation of BEC collective modes unique to hollow condensates.  Given the establishment of these techniques, bubble inflation (up to and beyond the few-mm scale) could drive new `model universe' experiments~\cite{Eckel:2018gv}, the fundamental limits of inflation adiabaticity and quantum behavior at dilute-BEC extremes could be explored---potentially with multiple species~\cite{wolf2021shell}---and bubble cooling and shaping techniques could be applied to spaceborne quantum sensing protocols~\cite{Lachmann.2021}.

\begin{acknowledgments}
The authors would like to thank the entire NASA/JPL Cold Atom Lab team for their guidance with these experiments.  We also thank Barry Garraway for helpful conversation over many years. This work was supported by the National Aeronautics and Space Administration through a contract with the Jet Propulsion Laboratory, California Institute of Technology.

\end{acknowledgments}






\section*{Author Information}
These authors contributed equally: Ryan A.~Carollo, David C.~Aveline
\subsection*{Contributions}
\noindent   R.A.C. designed experiments, guided data collection, and wrote analysis software.  D.C.A. conceived the study, designed experiments, guided data collection, operated the CAL instrument, provided scientific guidance, and prepared the manuscript.  B.R. performed modeling calculations, prepared the manuscript, and provided theory support.  S.V. and C.L.  conceived the study, guided model calculations, and provided scientific guidance and theory support.  J.D.M. prepared the manuscript and wrote analysis software.  E.R.E. and J.R.W. and R.J.T. operated the CAL instrument and guided data collection; R.J.T. and J.R.W. also provided guidance as CAL Project Scientists.  N.L. conceived the study, designed experiments, guided data collection, performed data analysis, and prepared the manuscript.  All authors read, edited and approved the final manuscript.


\clearpage

\clearpage
\section{Methods}
\noindent {\it User facility and sample preparation.---}
CAL is a multi-user research facility installed and operating aboard the ISS, where it has been running cold atom experiments on a daily basis since June of 2018. Remotely controlled from the Jet Propulsion Laboratory in Pasadena, CA, the instrument produces $^{87}$Rb Bose-Einstein condensates using an atom-chip device and carries a suite of tools that enable a variety of cold atom studies led by multiple investigators from around the world. The BEC production is based on laser-cooled atoms that are subsequently magnetically trapped and transported to an atom chip surface, where they are cooled via forced evaporation with rf radiation.   For use with the experiments associated with this work, the trapped sample was transported away from the chip in a process that reduced the needle-like aspect ratio of the original trap, and reduced the overall trap tightness.  Done rapidly and/or with simple linear ramps of trap parameters, this process could result in significant center-of-mass excitation rendering further work difficult; to ameliorate this, we applied custom expansion pathways based on formalism developed in Ref.~\cite{Sackett:2017gp} and observed that with sufficiently large overall expansion time residual trap motion could be reduced to $\lesssim$ 1 $\mu$m.   Deliberately rapid expansion pathways were used to excite measurable `sloshing' used to confirm our modeling of the chip trap system.  We denote the trapping frequencies as $\omega_i$ ($i=1,2,3$), where the principle oscillation direction associated with $\omega_1$ and $\omega_2$ lies in the $xy-$plane parallel to the atom chip, and that of $\omega_3$ lies in the $z$-direction (perpendicular to the chip).   Observations of $\omega_3$ yielded a best estimate of 2$\pi\times$100(1) Hz, consistent with our model prediction of 101 Hz; model predictions for the other frequencies were $\omega_1=2\pi\times$31 Hz, $\omega_2=2\pi\times$98 Hz.   Residual micromotion remaining from the sample expansion trajectory was estimated to be of characteristic amplitude $<$ 1 $\mu$m.  \\

\noindent {\it Imaging.---}
Measurement of the atom cloud density distribution is carried out using absorption imaging techniques with an optical path passing parallel to the chip surface (along the instrument's $y$-axis). The optical beam is approximately 10 mm in diameter and centered $\sim$4 mm below the chip surface, and is directed via collection optics into a CMOS camera with an associated magnification factor of 1.2. A small magnetic field is applied along the $y$-axis to enhance the absorption with this circularly polarized optical beam. The results reported here are collected using two 40-$\mu$s pulses separated by 53 ms, with the first pulse  containing atoms and the second serving as a reference.   The effective pixel size used for all analysis in this paper is $\ell=$ 4.52 $\mu$m.\\

\noindent {\it Thermodynamic model---}
Here we summarize our modeling approach for predicting thermometry in the shell geometries at hand. We consider two aspects, namely, i) the transition temperature $T_c$ at which we expect a fraction of condensate to appear for a given shell-shaped geometry and ii) the change in temperature of the ultracold atomic gas as the trapping potential evolves and gives rise to adiabatic expansion. 

At temperatures much larger than the single-particle energy level spacing, one can employ a semiclassical approximation~\cite{Pethick_2008,Pitaevskii_2016}. For a collection of noninteracting $^{87}\text{Rb}$ atoms, this amounts to using the energy relation ${\bf p}^2/2M + U({\bf r})$ where $M$, ${\bf p}$, and $U({\bf r})$ are the particle's mass, momentum, and confining potential, respectively. The validity of the semiclassical approximation for shell-shaped potentials is addressed in more detail in Ref. \cite{Rhyno.2021}; we find that it holds for a significant fraction of conditions considered here. In this scheme, standard thermodynamic sums over eigenstates of the Schr{\"o}dinger equation are replaced by integrals over position and momentum \cite{Pethick_2008,Pitaevskii_2016}. The momentum integrals can be performed analytically; for instance, one finds that the single-particle density of states takes the form
\begin{equation}\label{SC_DOS}
    \rho(\varepsilon)= \frac{2}{\sqrt{\pi}} \left( \frac{M}{2\pi\hbar^2} \right)^{3/2} \int d{\bf r} \, \theta(\varepsilon - U({\bf r})) \sqrt{\varepsilon - U({\bf r})}
    ,
\end{equation}
where the integration is over all space and $\theta(\cdot)$ denotes the Heaviside step function \cite{Bagnato:1987hd}. 
In order to carry out spatial integrals, we employ a numeric method. We create a spatial grid with typical lattice spacing 1 $\mu\text{m}$ and apply the numerically generated potential $U({\bf r})$.

As discussed in the main text, the dressed potentials of interest are characterized by a detuning frequency $\Delta$ which, when increased, inflates the size of the bubble. As a function of detuning, we use the semiclassical formalism to numerically compute both the transition temperature, $T_c(\Delta)$, and the temperature of the gas during adiabatic expansions, $T_{\rm bubble}(\Delta)$ given an initial temperature. 

In the thermodynamic limit, the transition temperature $T_c(\Delta)$ is found in the semiclassical approximation by setting the chemical potential equal to the minimum value of $U({\bf r})$ (which we set to zero here for convenience) and finding  the temperature that makes the number of excited particles equal to the total number of particles. Explicitly, for each dressed potential, we determine the temperature that satisfies the equation
\begin{equation}\label{SC_Tc}
    N = \int d\varepsilon \rho(\varepsilon) \frac{1}{e^{\varepsilon/k_B T_c}-1}
    .
\end{equation}
Alternately, by inserting Eq.~(\ref{SC_DOS}) and integrating over energy, this process could be performed using the following~\cite{Houbiers.1997}:
\begin{equation}\label{SC_Tc_space}
    N = \frac{1}{\Lambda^3_{\rm th}}\int d{\bf r}~g_{3/2}[e^{-U({\bf r}) /k_B T_c}]
    ,
\end{equation}
where $\Lambda_{\rm th}=\sqrt{2\pi\hbar^2/M k_B T}$ is the thermal de Broglie wavelength (evaluted at $T_c$) and $g_{s}[z] = \sum_{n=1}^\infty z^n / n^{s}$ is the Bose function.


Turning to adiabatic expansion modelling, we first fix the number of particles in our trap $N$ and the initial temperature of the system prior to expansion, i.e.~when the trap potential is at its lowest detuning frequency. Next, we find the entropy associated with this initial setup. This is done numerically by simultaneously solving the equations for particle number and entropy:
\begin{subequations}\label{SC_expansion}
    \begin{eqnarray}
        N &&=
        N_0 + \int d\varepsilon \rho(\varepsilon) f(\varepsilon)
        \label{SC_N},
        \\
        S &&= k_B
        \int d\varepsilon \rho(\varepsilon)
        \{ [1+f(\varepsilon)]\ln[1+f(\varepsilon)]
        \nonumber\\
        &&\qquad\qquad\qquad\qquad
        - f(\varepsilon) \ln f(\varepsilon)\}
        \label{SC_S},
    \end{eqnarray}
\end{subequations}
where $N_0$ is the number of condensed particles and $f(\varepsilon) = \{\exp[(\varepsilon-\mu)/k_B T] - 1\}^{-1}$ is the Bose-Einstein distribution function at temperature $T$ and chemical potential $\mu$.
Whereas below $T_c$ we have $\mu=0$, above $T_c$, where $N_0=0$, we must determine the chemical potential.
As in the calculation of $T_c$, one can carry out the energy integration to obtain convenient formulae for both the particle number and the entropy of a trapped Bose gas~\cite{Houbiers.1997}:
\begin{subequations}\label{SC_expansion_space}
    \begin{eqnarray}
        N &&= N_0 + \frac{1}{\Lambda^3_{\rm th}}\int d{\bf r}~g_{3/2}[z({\bf r})]
        \label{SC_N_space},
        \\
        S &&= \frac{k_B}{\Lambda^3_{\rm th}} \int d{\bf r}~\left\{\frac{5}{2} g_{5/2}[z({\bf r})] - g_{3/2}[z({\bf r})]\ln{z({\bf r})}  \right\}
        \label{SC_S_space},
        \quad
    \end{eqnarray}
\end{subequations}
where $z({\bf r})$ is the local fugacity $\exp{ [ ( \mu-U({\bf r}) ) / k_B T ] }$.


Once an initial entropy is known, the evolution of the temperature during expansion can be determined.  We increase $\Delta$ (considering a different dressed potential) and find the new temperature of the gas by simultaneously demanding both the semiclassical expressions for the total particle number and entropy above remain fixed. Holding the entropy constant is equivalent to demanding adiabaticity during the expansion. The results obtained using these methods are shown and discussed in the main text.  
The uncertainty bands on the theory curves for $T_c(\Delta)$ in Fig.~3 are approximately $\pm 10$ nK (originating in the spread of $N$ in a given dataset) and do not affect any interpretation of this work.

\noindent {\it Dressing Hamiltonian.---}  Below is the dressing Hamiltonian which is used for all our modeling, developed through application of a rotating frame and the rotating-wave approximation.

\begin{widetext}
\begin{equation}\label{ham}
{\mathcal H}=\left( \begin{matrix}
2\omega & \Omega/2 & 0 & 0 & 0 \\ 
\Omega/2 & \omega & \sqrt{\frac{3}{2}}\Omega/2  & 0 & 0 \\
0 &\sqrt{\frac{3}{2}}\Omega/2& 0 &\sqrt{\frac{3}{2}}\Omega/2& 0\\
0 & 0 & \sqrt{\frac{3}{2}}\Omega/2 & -\omega &\Omega/2 \\
0 & 0 & 0 & \Omega/2 & -2\omega \\
\end{matrix}\right) + {\mathcal H_{\rm Zeeman}({\bf r})}  
\end{equation}
\end{widetext}
where $ {\mathcal H_{\rm Zeeman}({\bf r})} $ is diagonal and represents the (exact) Zeeman shifts of the states in use, which for this work are those in the $^{87}$Rb upper hyperfine ground state manifold denoted by  $|F=2, m_F \rangle$, with $m_F$ taking values from -2 to 2.   Use of this Hamiltonian assumes that the coupling strength (set in this case by the Rabi frequency $\Omega$) is always and everywhere sufficiently large to ensure dressing adiabaticity, thus ensuring stability of atoms in a given $m'$ state, protected against `Landau-Zener' losses to lower-lying dressed spin states~\cite{Burrows:2017bw}.   Our typical operating parameter of $\Omega/2\pi =$6(1) kHz is consistent with lifetimes exceeding 150 ms, confirmed by hold-time measurements showing no significant loss.  These observations were performed with final rf-knife values of 5.00 MHz and 4.86 MHz and performed at apparatus detuning 160 kHz, or $\Delta\simeq +110$ kHz.           
  Given a driving frequency $\omega$ with coupling strength $\Omega({\bf r})$, we calculate the dressed potentials $U^*_{m'}({\bf r})$ as the spatially-dependent eigenvalues of  ${\mathcal H}$.  While not of specific interest in this work, the eigenvectors of ${\mathcal H}$ represent the decomposition of the dressed spin state of an atom at ${\bf r}$ into the lab-spin basis.   Accounting for terrestrial gravitational effects would require the addition of an $M g z$-like term to the Hamiltonian.  Inhomogeneities in the magnitude and direction of $\Omega$ result in effective gravitational tilts to the dressed potentials, discussed thoroughly in Ref.~\cite{Lundblad:2019ia}.

\noindent {\it Rabi calibration.---}
A crucial parameter in the observation and modeling of ultracold rf-dressed systems is the coupling strength $\Omega$.   In our case it is driven by the interaction between the atoms and a rf field originating in a nearby wire loop.  In general the coupling strength is state-dependent, spatially-dependent due to the inhomogeneous amplitude and direction of ${\bf B}_{\rm rf}$, and frequency-dependent due to the nature of the rf amplifier and coil design.  Nevertheless a single parameter is used as a basis for our modeling, with various inhomogeneities accounted for separately in the model.  We obtained a coupling parameter $\Omega/2\pi$---the Rabi frequency---using 5-level Rabi spectroscopy of the $F=2$ manifold. This was performed by preparing an ultracold sample in the $|2,2\rangle$ state in a trapping configuration somewhat relaxed from the initial tight trap.    We then switched off the trapping fields, maintaining a constant bias field of approximately 5.2 G.  After an rf pulse of 100 $\mu$s duration and variable frequency near 3.7 MHz, a Stern-Gerlach gradient was applied to separate differing spin components, followed by conventional absorption imaging.   The resulting 5-level Rabi spectra were fit using optimization routines (Mathematica) resulting in a conservative estimate of $\Omega/2\pi=$ 6(1) kHz (at this rf frequency) and an estimate of the constant bias field of 5.238(1) G, with uncertainty largely coming from shot-to-shot noise in the spin populations combined with imaging noise.  A separate effort taken by JPL/CAL researchers found a slightly higher Rabi frequency of $\simeq$8 kHz near 27 MHz, suggesting general broadband capability of the rf amplifier.   The data taken in this paper generally were taken with rf frequencies in the 2--3 MHz range, depending on initial and final shell inflation parameters.   \\   



\begin{figure}[t!]
\centering
  \includegraphics[width=3in]{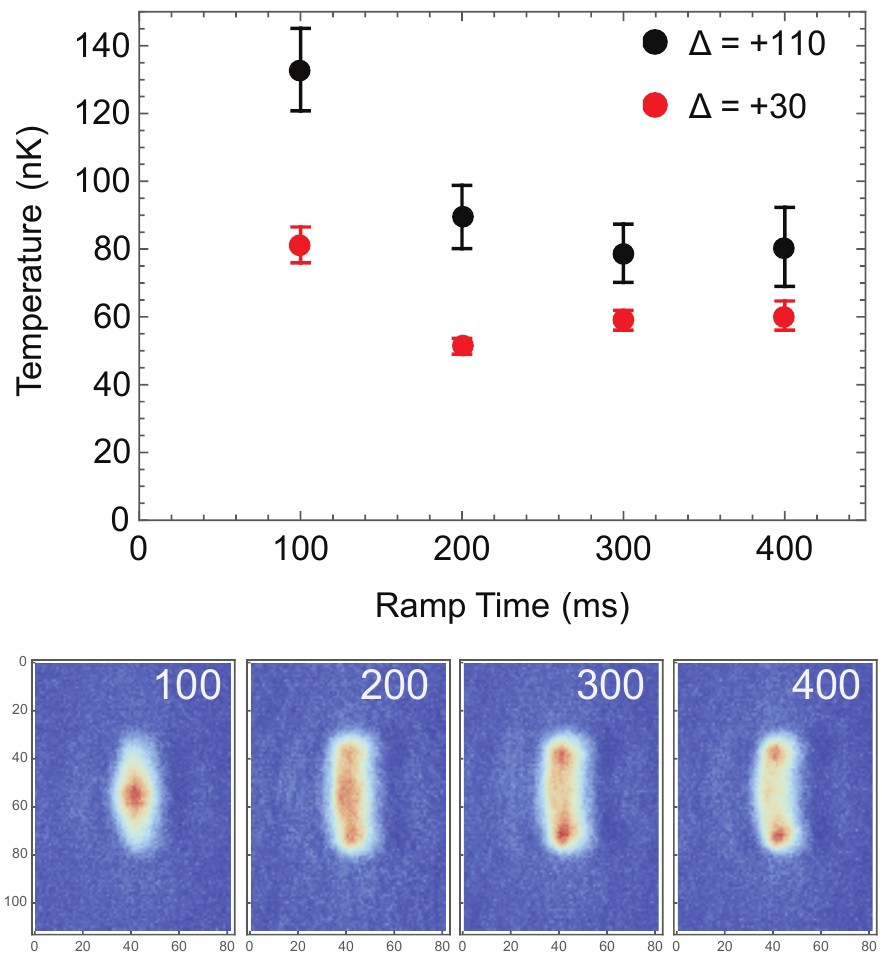}
  \caption{
Ramp time is varied 100--400 ms, with a 1000-point frequency ramp extending 200 kHz upward from an initial frequency of 2.05 MHz  $+ \Delta$, corresponding to variation in ramp speed 0.5--2.0 kHz/ms.   Bottom: absorption imaging of $\Delta=$ +30 kHz clouds associated with marked ramp times (associated with red points above).  For this dataset initial cloud temperature was set slightly below $T_c$, similar to that used in Fig.~3(d ). }     
\label{tdress}
\end{figure}

\noindent {\it Details of the rf ramp.---}
The rf radiation is generated by an AWG (National Instruments model PXI 5422), amplified, and emitted from a double loop (OD $\sim$ 10 mm) of copper wire located on the ambient side of the atom chip. This rf source is used for evaporative cooling and (specifically for this work) applied with low-to-high sweeps of frequency to dress the cold atom traps.  The rapidity of a frequency sweep is an influential parameter for maintaining adiabaticity in bubble inflation, both for the dressed potentials themselves (spin-following adiabaticity) and the mechanical adiabaticity associated with the deformation of the dressed trap potentials.   In Fig.~\ref{tdress} we show the results of thermometry performed on dressed clouds but with ramps of varying duration.   While no thermal difference is detected in this case beyond 100 ms ramp time, qualitative inspection of the dressed clouds suggests changes in density distributions as ramp time is varied.

\begin{figure}[t!]
\centering
  \includegraphics[width=3in]{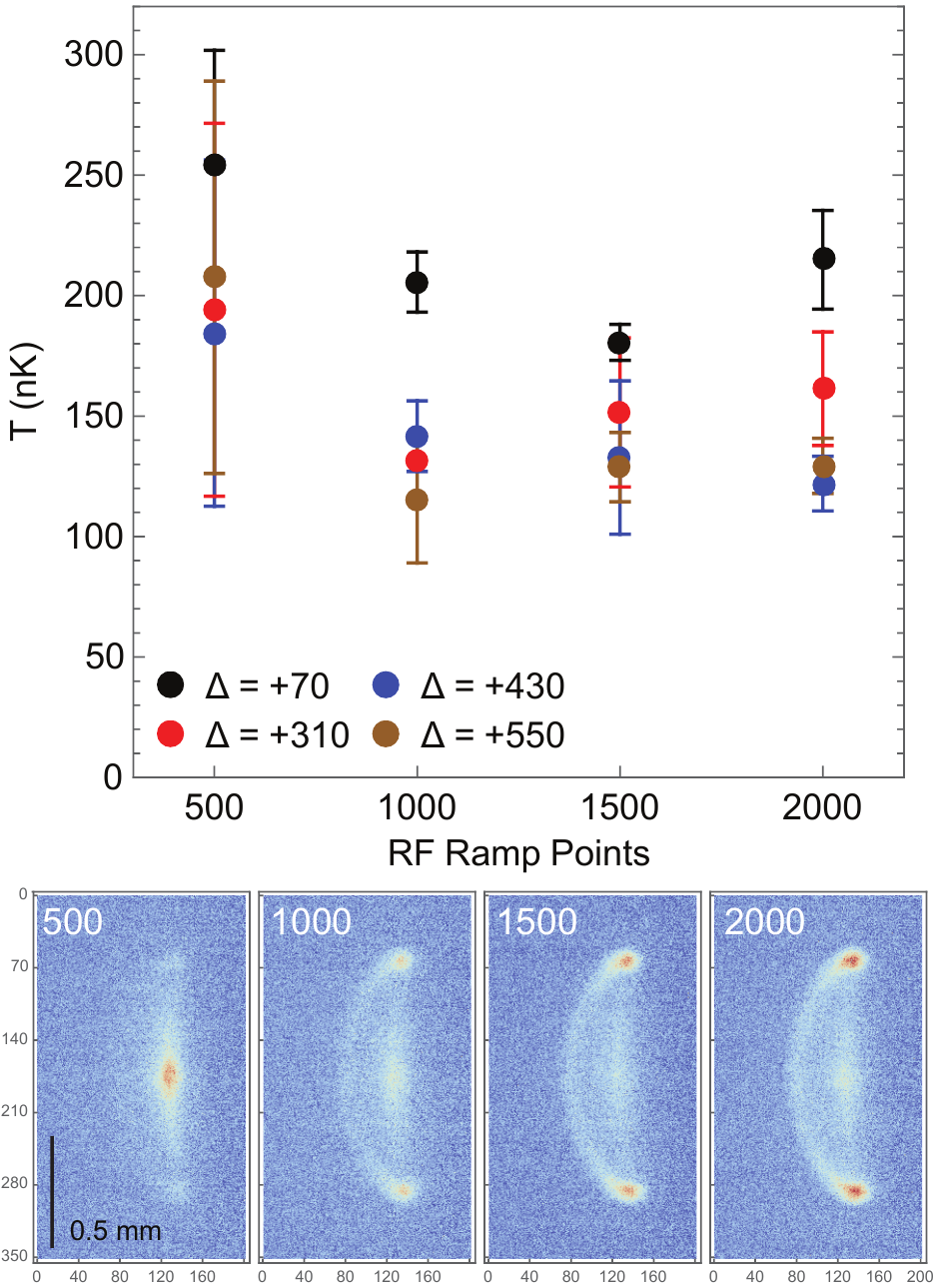}
  \caption{
Above: graining of the dressing ramp is varied, with resulting dressed-sample thermometry plotted as a function of the number of frequency steps. All dressing ramps extended 600 kHz upward from an initial frequency of 1.65 MHz + $\Delta$, over 400 ms (ramp speed 1.5 kHz/ms), thus varying the step size from 300--1200 Hz.  For this dataset initial cloud temperature was set significantly above $T_c$, similar to that used in Fig.~3(b).  Below: dressed ($\Delta$ = +550 kHz, i.e.~a ramp 2.2--2.8 MHz) clouds at short (2.6 ms) TOF  associated with each rf frequency step graining; note qualitative difference associated with 500-point graining.}     
\label{grain}
\end{figure}

As discussed in Ref.~\cite{Morizot:2008bq}, the step size of any noncontinuous frequency ramp impacts the adiabaticity of shell inflation; in Fig.~\ref{grain} we show the results of varying the number of discrete frequency steps in a given ramp of (relatively large) amplitude 600 kHz and duration 400 ms, with initial rf-knife set significantly above $T_c$ in order to yield sufficient absorption signal at this shell size.  The limit of graining (2000 points) was set by CAL hardware and operational parameters.  A clear increase in temperature (Fig.~\ref{grain} upper) was associated with sequences of 500 steps (1200 Hz / step) with inconclusive behavior for finer graining. Qualitative inspection of the associated dressed clouds suggested a change in density distribution associated with the  500-step ramps, as shown in Fig.~\ref{grain} (lower). 

As a result of these investigations, the  datasets of Fig.~3 in the main text are taken with dressing ramps of  300 kHz amplitude and  400 ms duration, with 1000 frequency steps (0.75 kHz / ms sweep rate, 300 Hz / step). \\

\clearpage

\end{document}